# Global Gridded Daily CO$_2$ Emissions

**Xinyu Dou[1], Yilong Wang[2], Philippe Ciais[3], Frédéric Chevallier[3], Steven J. Davis[4], Monica Crippa[5], Greet Janssens-Maenhout[5], Diego Guizzardi[5], Efisio Solazzo[5], Feifan Yan[6], Da Huo[1], Zheng Bo[7], Zhu Deng[1], Biqing Zhu[1], Hengqi Wang[1], Qiang Zhang[1], Pierre Gentine[8] and Zhu Liu[1]***

**Affiliations**
1. Department of Earth System Science, Tsinghua University, Beijing, China
2. Key Laboratory of Land Surface Pattern and Simulation, Institute of Geographical Sciences and Natural Resources Research, Chinese Academy of Sciences, Beijing, China
3. Laboratoire des Sciences du Climat et de l'Environnement, LSCE/IPSL, CEA-CNRS-UVSQ, Université Paris-Saclay, Gif-sur-Yvette, France
4. Department of Earth System Science, University of California, Irvine, CA, USA
5. European Commission, Joint Research Centre (JRC), Ispra, Italy
6. Key Laboratory of Marine Environment and Ecology, and Frontiers Science Center for Deep Ocean Multispheres and Earth System, Ministry of Education, Ocean University of China, Qingdao, China
7. Institute of Environment and Ecology, Tsinghua Shenzhen International Graduate School, Tsinghua University, Shenzhen, China
8. Department of Earth and Environmental Engineering, Columbia University, New York, NY, USA

*Correspondence: Zhu Liu; E-mail: zhuliu@tsinghua.edu.cn

**ABSTRACT:**
Precise and high-resolution carbon dioxide ($CO_2$) emission data is of great importance of achieving the carbon neutrality around the world. Here we present for the first time the near-real-time Global Gridded Daily $CO_2$ Emission Datasets (called GRACED) from fossil fuel and cement production with a global spatial-resolution of 0.1° by 0.1° and a temporal-resolution of 1-day. Gridded fossil emissions are computed for different sectors based on the daily national $CO_2$ emissions from near real time dataset (Carbon Monitor), the spatial patterns of point source emission dataset Global Carbon Grid (GID), Emission Database for Global Atmospheric Research (EDGAR) and spatiotemporal patters of satellite nitrogen dioxide ($NO_2$) retrievals. Our study on the global $CO_2$ emissions responds to the growing and urgent need for high-quality, fine-grained near-real-time $CO_2$ emissions estimates to support global emissions monitoring across various spatial scales. We show the spatial patterns of emission changes for power, industry, residential consumption, ground transportation, domestic and international aviation, and international shipping sectors between 2019 and 2020. This help us to give insights on the relative contributions of various sectors and provides a fast and fine-grained overview of where and when fossil $CO_2$ emissions have decreased and rebounded in response to emergencies (e.g. COVID-19) and other disturbances of human activities than any previously published dataset. As the world recovers from the pandemic and decarbonizes its energy systems, regular updates of this dataset will allow policymakers to more closely monitor the effectiveness of climate and energy policies and quickly adapt.

**INTRODUCTION**

Although human emissions of carbon dioxide ($CO_2$) to the atmosphere are the main cause of global climate change, detailed and spatially-explicit estimates of such emissions are updated infrequently, typically lagging emissions by at least a year. Yet a reliable, spatially-explicit and up-to-date dataset of fossil $CO_2$ emissions is becoming increasingly important with the rising ambition of climate policies and mitigation efforts. For example, such detailed data is necessary to link emissions to observable atmospheric concentration signals and constrain regional $CO_2$ fluxes, and can help decision makers to more quickly assess both the effectiveness of policies and local priorities for further mitigation[1,2].

Since the end of 2019, the COVID-19 pandemic has caused major disruptions of human activities and energy use. Governments around the world have imposed compulsory lockdowns that restrict in-person educational and commercial activities to reduce the spread of coronavirus. In turn, industries and factories reduced their activities and production, people's local and long distance mobility was reduced, and human activities were reduced on a large scale, resulting in a substantial decrease in fossil energy consumption and $CO_2$ emissions, albeit with large regional differences [2-4]. As lockdown restrictions have relaxed in many countries and economic activities have recovered in some sectors, the effect of the pandemic on $CO_2$ emissions has weakened, even during large "second waves" of cases. A timely and finely-gridded emissions dataset enables quantitative analysis of how temporal and spatial changes in $CO_2$ emissions in each country in response to emergencies (e.g. COVID-19) and other disturbances of human activities, and helps constraining predictions of future trends.

Existing datasets of global gridded (i.e. spatially-explicit) $CO_2$ emissions include the Open-source Data Inventory for Anthropogenic $CO_2$ (ODIAC) that distributes national emission totals estimated by the Carbon Dioxide Information Analysis Center (CDIAC) in space, using a combination of geospatial proxies such as satellite observations of nighttime lights and geolocations of major power plants(CARMA list): ODIAC provides maps of monthly $CO_2$ emissions on a 1 km grid for the period 2000 to 2019, as of today, including emissions from power plant, transportation, cement production/industrial facilities, and gas flares over land regions[5-7]. Similarly, the Community Emissions Data System (CEDS) uses data from a number of existing inventories to provide a monthly gridded dataset of all emission species for the Climate Model Inter-comparison Program (CMIP6) over the period 1750 to 2014 at a resolution of up to 0.1°, including sectors of energy transformation and extraction, industry, residential, commercial, transportation, agriculture, solvent production and application, waste, shipping and "other"[8-11]. Another prominent example is the Emission Database for Global Atmospheric Research (EDGAR). EDGAR estimates emissions based on national $CO_2$ emissions reported by the Global Carbon Project (GCP) and emission factors, broken down to IPCC-relevant source-sector levels. EDGAR uses spatial geospatial proxies such as point and line source locations at a 0.1°×0.1° resolution for the period 1970 to 2019, including sectors of agriculture, power, transport, residential, industry, manufacturing, and a number of others[12-15]. More recently, The Global Carbon Grid (http://gidmodel.org) establishes high-resolution maps of global $CO_2$ emissions from fossil fuel combustion and cement production based on a framework that integrates multiple data flows including point sources, country-level sectoral activities and emissions, and transport emissions and distributions. The Global Carbon Grid v1.0 provides global 0.1°×0.1° $CO_2$ emission maps of six source sectors, including power, industry, residential, transport, shipping, and aviation in 2019[16-18].

Even the most current of the gridded $CO_2$ emissions datasets described above lag emissions by a year or more and do not reflect sub-monthly temporal variations related to seasonality, weather, economic activities, or policies. Nassar et al. made a first attempt to further downscale these global datasets at the weekly and diurnal scale using static local temporal scaling factors[19]. However, during a normal year, day-to-day variations are due mainly to weather impacting heating/cooling demands of residences and commercial buildings and the generation of renewable energy, as well as weekends and holidays. Since the pandemic began in early 2020, though, daily variations have been perturbed by a multitude of other factors, including lockdowns, industrial production drops and recoveries, and changes in human behavior. Timely and quantitative analysis on the effects of these COVID-related changes on $CO_2$ emissions using tools such as inversion systems thus requires dynamic knowledge of global $CO_2$ emissions. It was this need for data that led to our development of the Carbon Monitor, a near-real-time daily dataset of global $CO_2$ emission at the national level[1,2]: https://carbonmonitor.org). Chevallier et al. disaggregated the daily national Carbon Monitor totals on a worldwide uniform grid using satellite retrievals of a pollutant co-emitted with $CO_2$ as a spatial proxy, without sectoral distinction[20]. Here, we considerably refine the approach by downscaling the daily national emissions from Carbon Monitor into a 0.1° × 0.1° grid for each of the seven sectors (power, industry, residential, ground transportation, domestic and international aviation, and international shipping), using sector-specific geospatial data from the Global Carbon Grid (GID) v1.0, the EDGARv5.0_FT2019 database for 2019, and $NO_2$ retrievals from the Tropospheric Monitoring Instrument (TROPOMI) aboard the Sentinel-5 Precursor satellite to provide a new spatially-explicit dataset of daily global $CO_2$ emissions covering the last two years since Jan 1st 2019, which we name GRACED. The first high-resolution near-real-time gridded fossil $CO_2$ emission emissions GRACED we presented will facilitate the adaptive management of emissions and the implementation of climate policy, which is of great importance of achieving the carbon neutrality around the world.

## RESULTS

### Quarterly mean emissions

The global daily average emissions from all sectors of GRACED in 2020 is shown in **Figure 1**. GRACED demonstrates fine-grained emission differences produced by the allocation of emissions at the sub-national level. Emissions are shown at a common 0.1×0.1° resolution. In the figure, the five major global regions, including U.S. mainland, Europe, South East Asia, East Africa and Middle East, South America, are enlarged and displayed. It is showed that the spatial distribution characteristics of daily average emissions throughout 2020 are clustered, concentrated in areas such as eastern U.S., western Europe, southeastern China, South Korea, Japan, and India, with megacities as hotspots. The daily average total emissions in 2020 are approximately 3821 kgC per day per cell. The cell with the maximum emission value is 41320 tC per day per cell.

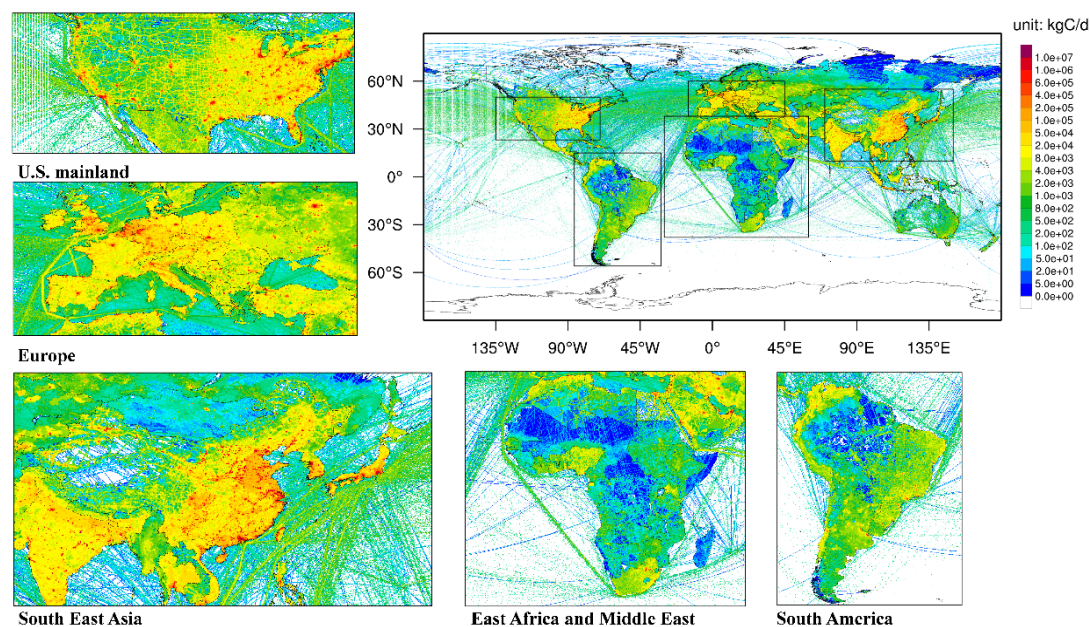

**Figure 1. The fossil fuel and cement $CO_2$ emissions distributions of GRACED in 2020. The value is given in the unit of kg of carbon per day per cell.**

We also calculate quarterly daily average total(**Figure S1**) and sectoral(**Figure S2**) emissions of 2020. We define January, February, and March as the first quarter, and then use this to introduce other months included in other quarters. The average total emission in the first quarter, is the highest with 3969 kgC per day per cell, and the average total emission quarter is the lowest in the second quarter, with 3381 kgC per day per cell.

As about 90% of the world's population is located in the northern hemisphere, the level of human activities in the northern hemisphere dominates the values of global emissions. The residential consumption sector and the aviation sectors, generate the most emissions in the fourth quarter, resulting in the highest average total emissions in the first quarter. Except for the residential consumption, industrial and international shipping sectors, the average lowest emissions from the other sectors all appear in the second quarter, which dominates the results of the lowest average total emissions in the second quarter.

**Range of daily emission variations**

The emission variations in different regions of the world in 2020 are shown in **Figure 2**. In 2020, the global grid average variation value is 4417 kgC/d. From a regional perspective, Europe, U.S., China, Southeast Asian countries, India, Japan, South Korea, etc. all have areas with large emission variation values (shown as red areas), and these areas are mainly distributed in economically developed areas, such as the Beijing-Tianjin-Hebei circle in China, the Yangtze River Delta, the Pearl River Delta of China, and California, Utah, and the eastern coastal areas of the U.S.. The emissions in Africa and South America have smaller variation value in 2020 (shown as blue areas). In 2019, the global average variation value of grids is 2930 kgC/d, which is smaller than that in 2020(**Figure S3**).

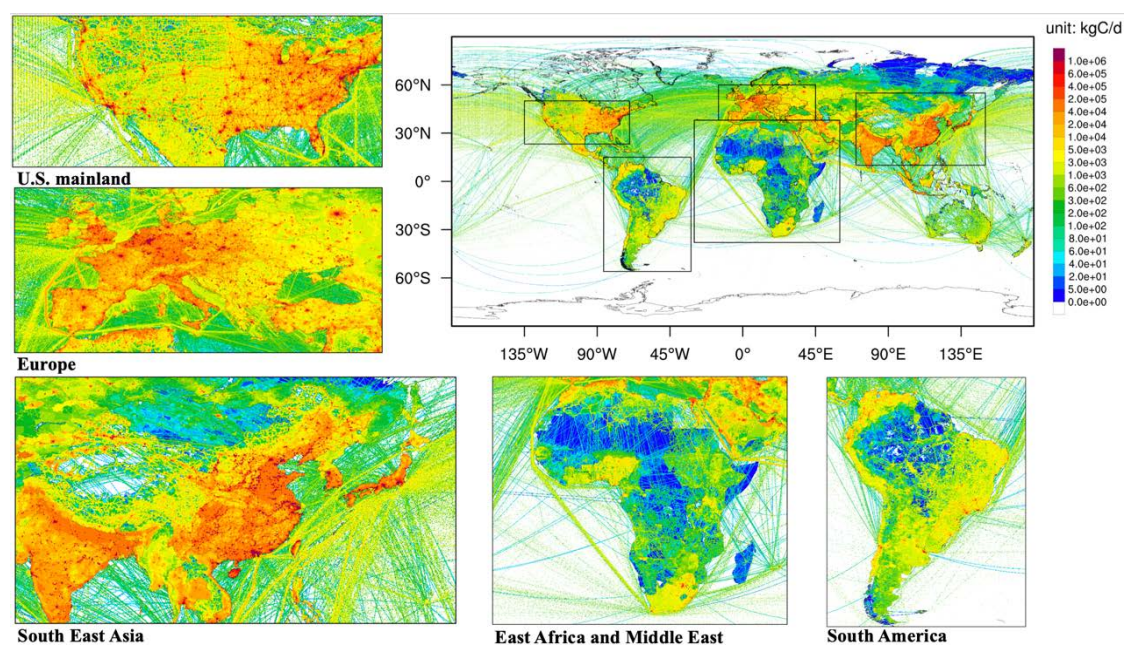

**Figure 2. The range value of daily variations of total emissions in 2020.**

A low standard deviation of all daily values in a quarter indicates that the emission values tend to be close to the mean (also called the expected value) of the set, while a high standard deviation indicates that the emission values are spread out over a wider variation.

From a quarterly point of view, on average, the distribution of global emission values in the first quarter of 2020 is the most heterogenous, with an average standard deviation value of 811 kgC per day per cell (**Figure S4**). The distribution of global emission values is the most homogenous in the third quarter, with an average standard deviation value of 625 kgC per day per cell. Besides, there are a standard deviation value of 791 kgC per day per cell in the fourth quarter and 634 kgC per day per cell in the second quarter.

**Difference between weekend and weekday emissions**

We then investigate the difference between weekend emissions and weekday emissions in **Figure 3.** It can be seen that on average, the global $CO_2$ emissions on weekends are generally less than the $CO_2$ emissions on weekdays. The global average of this difference is -248 kgC per day per grid. It can be further seen that the more developed regions have more significant differences between weekdays and weekends than the less developed regions (shown as the dark blue areas in the figure). Moreover, the spatial distribution characteristics of this difference showed an obvious linear relationship with the ground transportation sector's emission. It indicates that the reduction of human driving activities on weekends has a very important impact on the reduction of weekend emissions.

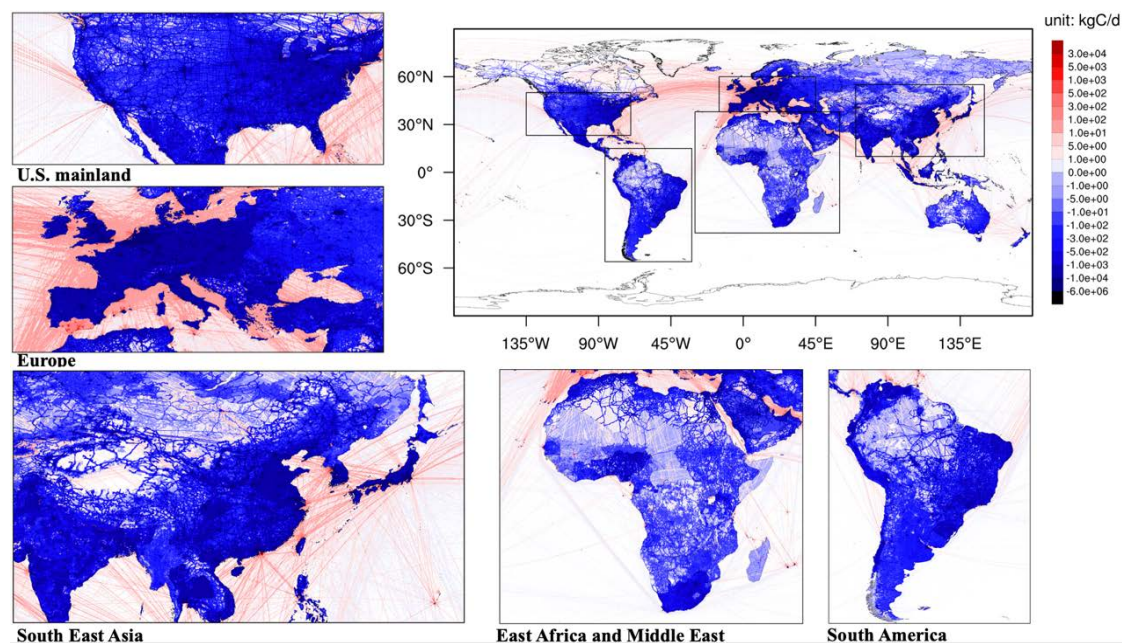

**Figure 3. Map of weekend minus weekday emissions in 2020.**

In 2019, on average, the global carbon emission on weekends is generally less than the emission on weekdays (**Figure S5**). The average value of this difference is -303 kgC per day per grid globally, which is higher compared with 2020. This is mainly because, affected by the COVID-19 in 2020, general human travel has generally reduced under the lockdown measures, at the same time, the implementation of the home office policy has weakened commuting travel during weekdays, making the difference in emissions between weekends and weekdays in 2020 less significant.

**Emission changes due to COVID-19**

Affected by the COVID-19 pandemics in 2020, compared to 2019, total emissions have generally declined worldwide (**Figure 4**). There were however a few regions experiencing an emission increase such as the eastern U.S., the United Kingdom, some areas of Europe, southeastern India, Japan's some provinces, and central and western China.

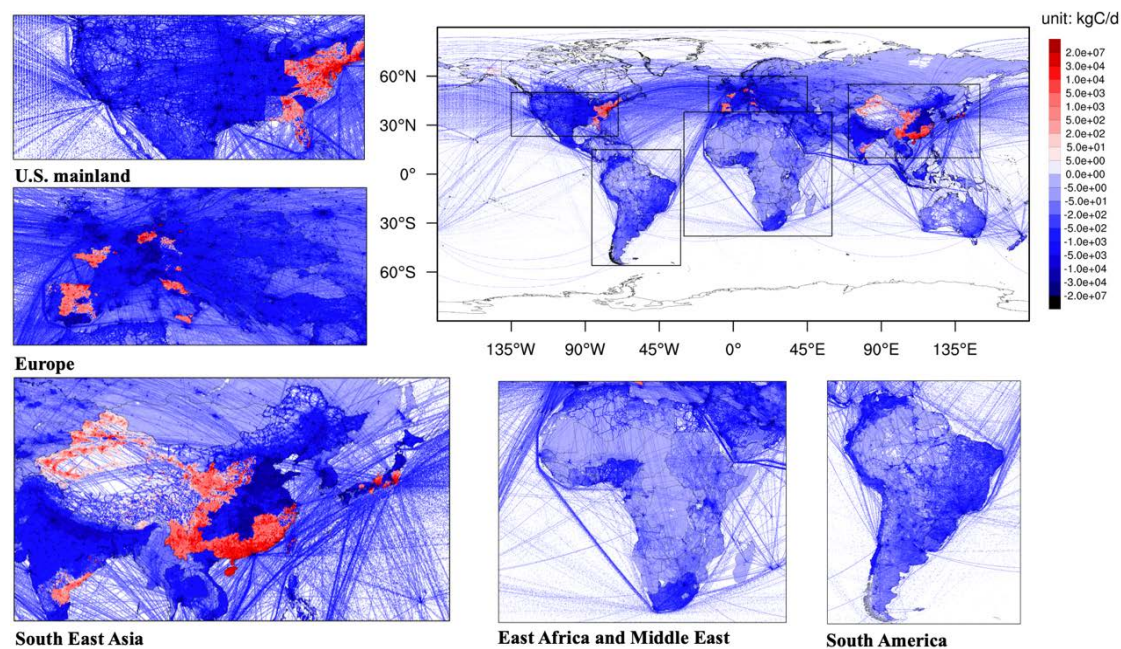

**Figure 4. Difference in daily average $CO_2$ emissions between 2020 and 2019 (2020 minus 2019).**

The dates with the maximum reduction and the maximum rebound in different regions in 2020 compared to 2019 reflect the sequence of the significant reduction in human activities caused by the severe impact of the COVID-19, **Figure 5a**. In this study, we define the date with the maximum rebound as the date appears the biggest increase in emissions in 2020 compared to 2019. The dates with the maximum rebound in different regions in 2020 compared to 2019 reflect the sequence of the largest economic recovery in the later period, **Figure 5b**.

In **Figure 5a**, obviously, some international aviation and international shipping were the first to be hit, which is shown in dark blue lines. From a national perspective, China's largest decline in 2020 appeared earliest compared to other countries, and time wise is closely related to China's first hit by COVID-19. While, the U. S., Spain and other countries experienced the largest emission reduction later. Most regions of India and Japan experienced this situation soon after. This may be mainly related to the late impact of the first wave of the COVID-19 in these countries and the more severe impact of the COVID-19's second wave in the later period. Furthermore, it can be seen that even within the same country, there were still large differences in the lockdown induced time of the largest reduction. For example, the time of the largest reduction in emissions due to the severe impact of the epidemic in different states in the U. S. varied.

Judging from the date with the largest rebound (**Figure 5b**), China, Russia, Myanmar, some European countries such as Netherlands, Poland, Italy, and some other countries experienced the largest rebound later, while in India, some states in the U.S. and some European countries such as Spain, Belarus and Ukraine, the biggest rebound occurred earlier.

Moreover, comparing the largest reduction timing, it can be seen that China experienced the largest reduction earlier and the largest rebound later, while India saw the largest reduction later and the largest rebound earlier. For countries in North America, Europe, Africa, and Southeast Asia, the time of the largest reduction was generally late, but there are significant differences in the time of the largest rebound.

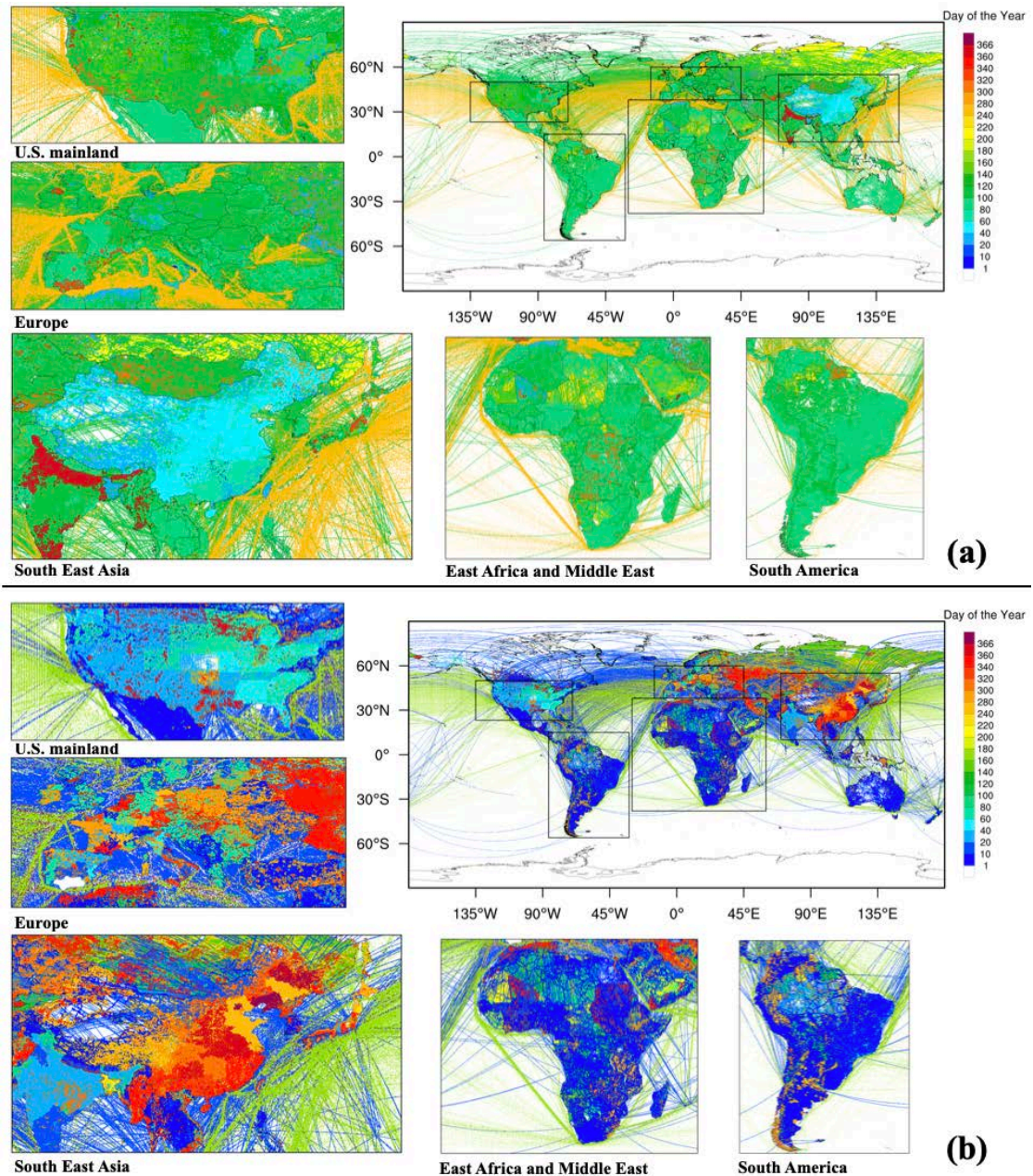

**Figure 5. The day of the year with (a) the maximum reduction, and (b) the maximum rebound of each grid in 2020 compared with 2019.**

**Sectoral emissions share**

Different sectors exhibit various spatial patterns. The emissions shares of various sectors in 2020 are shown in **Figure 6a-g**.

As shown in **Figure 6a**, the emissions share from power generation is generally high in the total emissions of the grid to which it belongs. Particularly, the power emissions share in the Democratic Republic of Congo in Africa is not high. This is mainly because this country is rich in hydropower resources, with hydropower accounting for almost 90%. For the industry sector, its emission share in developed countries such as the U.S., Australia, and Europe is generally low across the country (shown as the light areas in **Figure 6b**), while in China, India, Russia, Southeast Asia, and Africa, the emission share is relatively high, which is represented by the dark areas in **Figure 6b**. It reveals that the industrial activities of these countries still occupy an important position in their national economic activities in 2020. The development pattern of the residential consumption sector is quite different worldwide, and even within the same country, the emissions share of residential consumption sector varies significantly. This is mainly caused by the difference in regional population distribution and activity levels (**Figure 6c**). As shown in **Figure 6d**, ground transportation emissions account for a relatively high proportion of the total emissions worldwide. While in China, India, and Russia, the proportion to the total emission is not as high. This is mainly because their industry emissions are relatively high, making the share of ground transport emissions relatively low. As shown in **Figure 6e**, the emissions share of the international aviation sector is low in most of the land area, except in northern Africa, where the total emissions are low due to economic underdevelopment, making aviation routes through these regions accounted for a relatively high proportion. As international shipping emissions are only distributed in the marine area and only overlap with the spatial distribution of the international aviation sector, its emissions share is generally high (shown as dark areas in **Figure 6f**). For the domestic aviation sector in **Figure 6g,** its emission share varied significantly over the world. Its share is low in southeastern China, western Europe, and some states in the United States, while it is high in the western and central United States, Canada, Russia, and Australia.

At the grid level, changes in sector share between 2020 and 2019 are also observed (**Figure 6h-n**). The large-scale light areas in **Figure 6h** show that, compared with 2019, emissions share of the power sector in most regions throughout 2020 have declined, while this share in Europe and parts of China have increased. In **Figure 6i**, changes in emissions share from industry showed a decline in Europe and India, but shown an increase in most other regions of the world. At the same time, the changes in $CO_2$ emissions share from the residential consumption sector is more uniform in **Figure 6j**, with almost all regions increasing from 2019 to 2020, which is not only due to changes in population distribution and changes in residential emissions, but also the reduction in the share of other sectors impacted by the COVID-19 pandemics. The areas where ground transport emissions share increases are concentrated in western Europe, Russia and Middle East, while the decline in the share mainly occurs in southeastern China, U.S., and most of other regions (**Figure 6k**). The changes in emissions share of international aviation sector are quite uniform, with almost all regions decreasing from 2019 to 2020 (**Figure 6l**). In contrast, the international shipping sector shows a developing pattern. Compared with 2019, international shipping emissions share in 2020 in almost all regions showed an increase (**Figure 6m**). For the domestic aviation sector, its emissions share in 2020 in almost all regions showed an increase compared with 2019(**Figure 6n**).

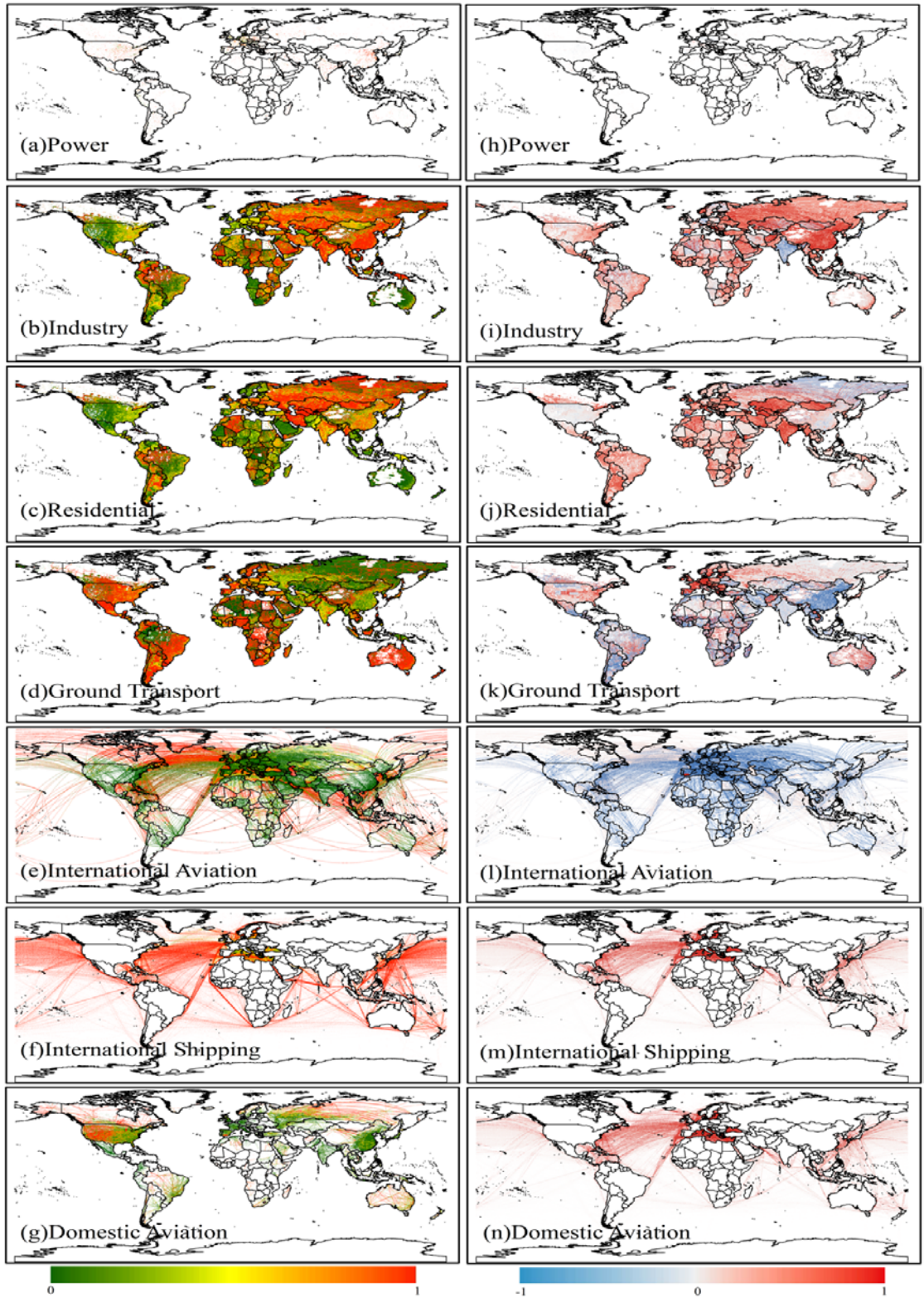

**Figure 6**. Sector share of $CO_2$ emissions in 2020(a-g). Difference in sector share of $CO_2$ emissions between 2020 and 2019(2020 minus 2019)(h-n).

**Uncertainty analysis**

The uncertainties are from Carbon Monitor, GID, and EDGAR dataset we used.

The uncertainty analysis of Carbon Monitor was presented in our related paper recently published at Nature Communications[2]. We followed the 2006 IPCC Guidelines for National Greenhouse Gas Inventories to conduct an uncertainty analysis of the data. First, the uncertainties were calculated for each sector. The uncertainties ranges of the Power, Ground transport, Industry, Residential, Aviation, International shipping sector are ±14.0%, ±9.3%, ±36.0%, ±40.0%, ±10.2%, and ±13.0%, respectively. The uncertainty in the emission projection for 2019 is estimated as 2.2% by combining the reported uncertainty of the projected growth rates and the EDGAR estimates in 2018.Then, we combine all the uncertainties by following the error propagation equation from the IPCC. Eq. 5 and calculated that the overall uncertainty range of Carbon Monitor is ±7.2%.

As for GID and EDGAR, uncertainty is introduced in the magnitude of national-level total emissions, the magnitude and location of large point sources, the magnitude and distribution of non-point sources, and from the use of proxy data to characterize emissions. As pointed out by Hogue et al., the largest uncertainty contribution in gridded emission data sets comes from how well the distribution of the proxy used for spatial disaggregation represents the distribution of emissions[21]. So for the gridded data from GID and EDGAR used in this research, the largest contribution to uncertainty comes from the spatial disaggregation process of national-level emissions and the accuracy of the spatial proxy parameters. The subtraction of the sum of all precise point sources with little uncertainty from the national total of a specific sector leaves a remaining emission composed of smaller sources. Due to lack of information, the remaining emission is usually allocated based on e.g. a population density proxy. The uncertainties of the point sources and the remaining smaller sources are greatly different, which are larger than the uncertainty of the national total of a specific sector. The representative information about the selected characteristic parameters of the point sources is most critical and needs to be evaluated by measurements (such as on-site atmospheric measurement of $CO_2$ emission pollutants), but in-depth analysis beyond the scope of this paper would be required.

**DISCUSSION**

This research presents for the first time the near-real-time high-resolution gridded fossil $CO_2$ emissions from fossil fuel and cement production, which is based on the Carbon Monitor project[2], In this work, we developed a near real-time global gridded emission dataset called Carbon Monitor-G to provide a high-quality, fine-grained dataset since January 1, 2019. This dataset is a daily gridded map with a spatial resolution of 0.1°×0.1°. One of the advantages of GRACED is that it can support global near real-time carbon emission monitoring on various fine spatial scales (such as cities) at sub-national level, which can further improve our understanding of the spatio-temporal variability in emissions and human activities. Especially during the COVID-19 pandemics, it helps to more quickly grasp the changes in human activities under the implementation of different lockdown strategies in region-specific, sub-national or urban case studies. Through the time series of GRACED, we provide important input for the analysis of emission trends during the COVID-19 pandemics, which will help to carry out more local and adaptive management of climate change mitigation in post-COVID era, and is crucial for the realization of zero carbon emission target globally by 2050.

We found that carbon emissions are mainly concentrated in eastern U.S., western Europe, southeastern China, South Korea, Japan, and India spatially. A sharp decline of $CO_2$ emissions in 2020 was identified in the central and eastern U.S., the United Kingdom, France, Germany in Europe, India, Japan, South Korea and eastern China. Various sectors show different spatial distribution characteristics, which is mainly explained by the emission sources. We also investigated the spatial distribution of different sectors' emission shares. Overall, the power sector has the highest share for all regions/countries, while international aviation sector has the lowest one. However, the second contributing sector differs for different regions. In addition to the power sector, the emissions in China and India are dominated by industry sector, while ground transport dominates the emissions in U.S. and Europe in 2020. Finally, the changes in the emission shares of various sectors compared with 2019 were evaluated in this research, aiming to reveal sector-specific spatial development patterns impacted by the COVID-19 pandemics.

In general, the current statistical data cannot fully grasp the fine-grained dynamics of $CO_2$ emissions under the COVID-19 pandemics, and further monitoring, observation and data collecting are urgently needed. The ability of near real-time fine-grained monitoring of daily emission trends we demonstrate here helps to take timely local actions in regional, sub-national or urban areas, and have policy implications for local climate change mitigation and earth system management.

GRACED provides the first global near real-time gridded carbon emissions data. This globality and timeliness comes at the expense of reduced accuracy due to near real-time spatial allocation information. Therefore, it is recommended that potential users of GRACED carefully consider these limitations when using this dataset. Inevitably, with the updated version of proxy data, the accuracy of emission spatial allocation in future versions of GRACED can be further improved. With Carbon Monitor national-level data updated in near real time, we should be able to continue to produce updated future versions of GRACED products within the same model framework.

**MATERIALS AND METHODS**

**Datasets used in the study**

(1) A near-real-time daily dataset of global sectoral $CO_2$ emission from fossil fuel and cement production at national level since January 1, 2019 published as Carbon Monitor (data available at https://carbonmonitor.org/)[2]. (2) Global sectoral $CO_2$ emissions annual data with high-resolution of 0.1° in 2019 based on a framework that integrates multiple data flows including point sources, country-level sectoral activities and emissions, and transport emissions and distributions released by the Global Carbon Grid (http://gidmodel.org)[16-18,22-24]. (3) Global monthly gridded emissions at a 0.1°×0.1° resolution in 2019 defined for a large number of IPCC sub-sectors provided by the Emission Database for Global Atmospheric Research (EDGAR) (https://edgar.jrc.ec.europa.eu/overview.php?v=50_GHG) [14,25]. (4) Daily $NO_2$ TCVD retrievals data in 2019 and 2020 from the Tropospheric Monitoring Instrument (TROPOMI) on board the Sentinel-5 Precursor satellite, launched in October 2017.

The ground resolution of the TROPOMI $NO_2$ retrievals was $7 \times 3.5$ km$^2$ at nadir until 5 August 2019 and has been $5.5 \times 3.5$ km$^2$ afterwards, achieving near-global coverage in one day. Standard retrievals from the official offline processing with a quality assurance value greater than 0.75 were aggregated to daily time scale on a regular 0.1° × 0.1° global grid and averaged over 14-day averaging periods in order to reduce the retrieval noise and limit gaps in the retrievals.

**Spatial gridding methodology**

**Grouping the GID and EDGAR sectors into Carbon Monitor categories**

Firstly, we link the Carbon Monitor emission sectors to GID and EDGAR sectors according to **Table 1**. We consider that GID has the highest accuracy in source location and we rely on this database as much as possible. However, for the domestic and international aviation sector, and the international shipping sector, GID does not distinguish between related domestic and international sub-sectors: we therefore directly use EDGAR's monthly spatial patterns for the spatial distribution in these sectors.

**Table 1. Correspondence between Carbon Monitor categories and GID/EDGAR categories.**

| No. | Carbon Monitor categories | GID categories | EDGAR categories |
|---|---|---|---|
| 1 | Power | power | |
| 2 | Industry | industry | |
| 3 | Residential consumption | resident | |
| 4 | Ground transport | transport | |

| | | |
|---|---|---|
| 5 | International aviation | AIR Bunker oil for international transport |
| 6 | International shipping | SEA Bunker oil for international transport |
| 7 | Domestic aviation | AIR Bunker oil for domestic transport |

Secondly, we do spatially gridding procedure. We use the global annual spatial patterns of $CO_2$ emission from the GID sub-sectors and global monthly $CO_2$ emission spatial patterns from EDGAR sub-sectors for the year 2019 for spatially downscaling Carbon Monitor daily national-level emissions. We assume that the spatial pattern of emissions remained unchanged after the last year of GID and EDGAR (2019). The validity of this assumption will depend on the country and on the time horizon for the adjustment. While, the sub-national emission may change rapidly within a country from 2019 to 2020 as there was a great difference in the timing and degree of the impact of COVID-19 in various regions. Therefore, for large countries, we use sub-national proxy based on TROPOMI $NO_2$ retrievals data to allocate national Carbon Emission totals into regional totals, before doing a second down-scaling at 0.1° based on the GID and EDGAR spatial patterns. The analysis can be updated consistently with the latest high-resolution emission maps and other spatial proxies for each year.

The spatial disaggregation framework used in the GRACED is shown in **Figure 7**. It is a top-down methodology that allocates Carbon Monitor national-level daily emissions to finer grid cells using spatial patterns provided by GID and EDGAR and sub-national proxy based on TROPOMI $NO_2$ retrievals.

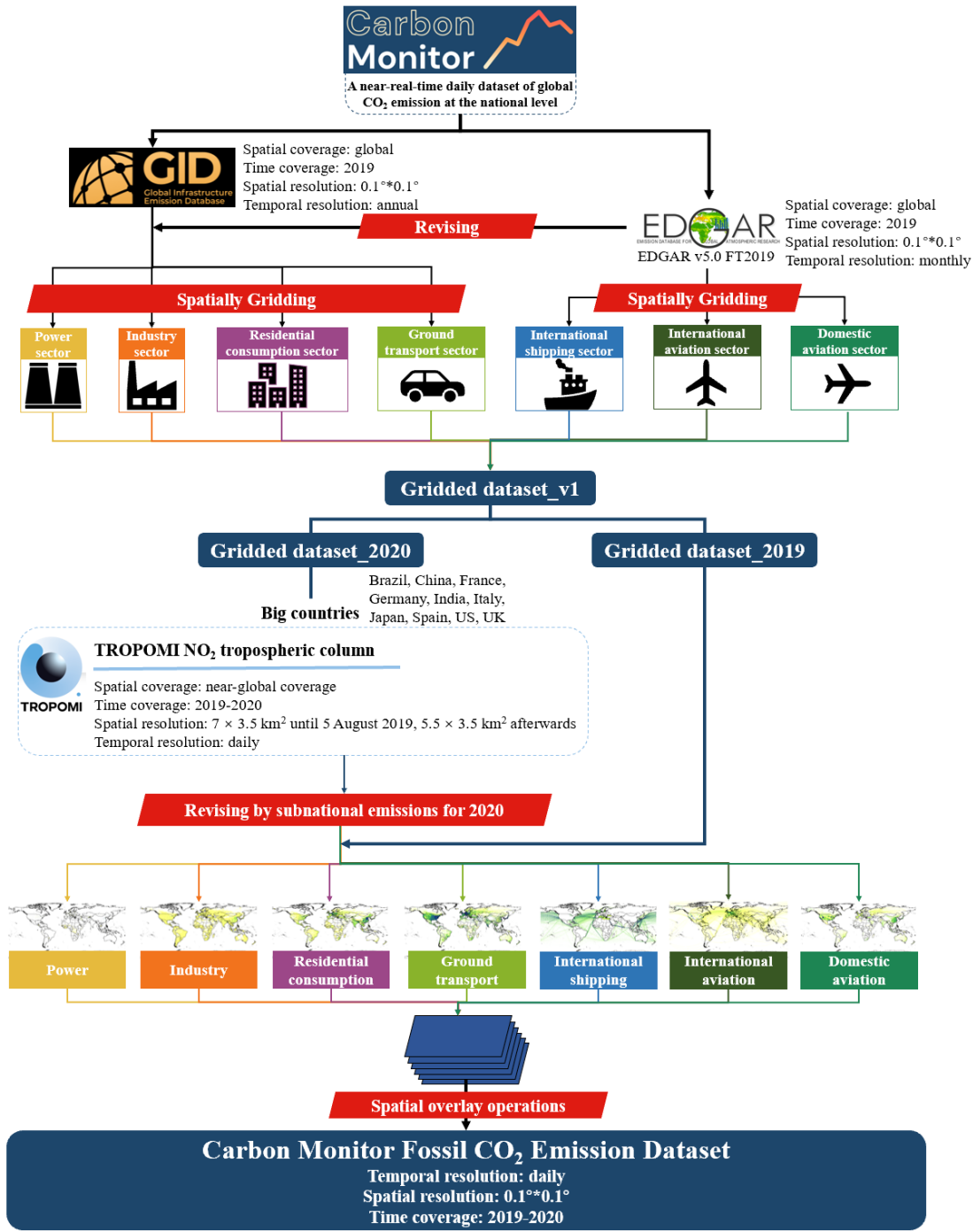

**Figure 7.** The framework of top-down spatially gridding methodology.

The detailed process of the model is presented as follow:

(1) Firstly, we use the spatial patterns provided by GID to allocate the national-level emissions of the four sectors of Carbon Monitor (Power, Industry, Residential consumption, Ground transport sector, see Table 1) from Carbon Monitor to obtain the daily gridded emissions under GID's annual spatial patterns. We then integrate the monthly spatial patterns of EDGAR for further correction, to correct the daily gridded emission based on GID's annual spatial patterns before under the monthly spatial patterns. For the domestic, international aviation sector, and international shipping sector, as GID does not distinguish between related domestic and international sub-sectors compared to EDGAR, we directly use EDGAR's monthly spatial patterns for distribution. The first version value of emission $Emi\_v1_{g,d,s}$ for grid $g$, date $d$ and sector $s$ as:

$$Emi\_v1_{g,d,s1} = CM_{country,d,s1} * \frac{GID_{g,s1}}{\sum_{i=1}^{n} GID_{i,s1}} * \frac{EDGAR_{g,m,s1}}{\sum_{j=1}^{12} EDGAR_{g,j,s1}} * 12 \quad (1)$$

$$Emi\_v1_{g,d,s2} = CM_{country,d,s2} * \frac{EDGAR_{g,m,s2}}{\sum_{i=1}^{n} EDGAR_{i,m,s2}} \quad (2)$$

Where $CM_{country,d,s}$ means the value of Carbon Monitor for country *country* which grid point $g$ belongs to, day $d$ and sector $s$. *s1* belongs to one of the sectors Power, Industry, Residential consumption, and Ground transport. *s2* includes International aviation, Domestic aviation, and International shipping. $GID_{g,s1}$ means the value of GID gridded $CO_2$ emission for grid point $g$ and sector *s1*. $n$ is the total number of grid points within this country. $EDGAR_{g,m,s}$ means the value of EDGAR for grid point $g$, month $m$ which date $d$ belongs to and sector $s$. $j$ is the index of a month.

(2) For big countries, subnational emission patterns can vary significantly from one year to the next. This was particularly obvious in 2020 with regional variations in the COVID-19 crisis, for instance between eastern and western US, or between eastern and western China. Capturing those sub-national emission changes is important for having a competitive product, and is not addressed by eq. (1) and eq. (2) which use a climatological emission pattern. It is reported that the global changes in emissions is also consistent with global changes in the NO2 inventory from satellite data[4]. Therefore, we assume that the sub-national emission changes follow the pattern of the differences in $NO_2$ column concentration between 2020 and 2019. In detail, we calculate an index R of each province, which is the averaged $NO_2$ concentration of each province, according to TROPOMI $NO_2$ retrievals data in year *y*.

$$R_{p,y} = NO_{2_{p,y}} \quad (3)$$

Where *p* represents province (state); *y* represents the year. $NO_{2_{p,y}}$ is the satellite $NO_2$ concentration averaged temporally over rolling 14 day-period in year *y* for province *p*(as explained above) and spatially over the 5% grid points within each province(state) that have the largest $NO_2$ average over the year. The choice of the 5% largest values allows extracting clear patterns very close to emission location. In the following step, we remove any negative $NO_2$ value for the 5% grid points over the year 2019 and 2020 that may be generated and attribute the mass gain to the other 5% pixels. Last, we calculate index R of each province in 2019(2020) according to TROPOMI $NO_2$ retrievals data.

Then we generate $CM_{p,d,s1,2020}$, the daily provincial emission in day $d$ and for sector $s1$ adjusted by the TROPOMI $NO_2$ retrievals in day $d$ and for sector $s$ in 2020 that matches the daily national total from Carbon Monitor following Eq. 5.

$$CM_{p,d,s1,2020} = \frac{CM_{p,d,s1,2019} * \frac{R_{p,2020}}{R_{p,2019}}}{\sum_{p=1}^{np} CM_{p,d,s1,2019} * \frac{R_{p,2020}}{R_{p,2019}}} \times CM_{country,d,s1,2020} \quad (5)$$

Where $CM_{p,d,s1,2019}$ means the first version of the emission value of a province in day $d$ and for sector $s1$ in 2019. $np$ is the number of provinces of the country. In detail, firstly, we calculate the ratio of change in the R index in 2020 compared to 2019, which is $\frac{R_{p,2020}}{R_{p,2019}}$. Secondly, multiply the provincial emission value aggregated from our first version dataset for 2019, $CM_{p,d,s,2019}$, to update the provincial emission value for 2020. Last, divide the updated provincial emission value by the sum of the updated provincial emission value $\sum_{p=1}^{np} CM_{p,d,s1,2019} * \frac{R_{p,2020}}{R_{p,2019}}$ in 2020 to do the normalization processing in the Eq.5. So, the sum of the updated provincial emissions within a country can be consistent with the national-level emission value from Carbon Monitor in 2020 after multiplying the national-level emission $CM_{country,d,s1,2020}$ from Carbon Monitor.

Then, based on the updated provincial emission $CM_{p,d,s1,2020}$ in 2020, we use GID and EDGAR data as the spatial patterns to distribute the emission data of each province for large countries to obtain our final version gridded emission value $Emi\_v2_{g,d,s}$.

$$Emi\_v2_{g,d,s_1} = CM_{p,d,s1,2020} * \frac{GID_{g,s1}}{\sum_{i=1}^{n} GID_{i,s1}} * \frac{EDGAR_{g,m,s1}}{\sum_{j=1}^{12} EDGAR_{g,j,s1}} * 12 \quad (6)$$

Where $n$ means the total number of grids within this province.

After revising the gridded emissions for large countries Brazil, China, France, Germany, India, Italy, Japan, Spain, US, and UK in 2020, GRACED is finally generated.

**DESCRIPTION OF SUPPLEMENTAL INFORMATION**

Supplemental Information includes four material sections, one text section, seven figures, one table and supplemental references.

## DECLARATION OF INTERESTS

The authors declare no competing interests.

## ACKNOWLEDGMENTS

Authors acknowledge the National Natural Science Foundation of China (grant 71874097 and 41921005), Beijing Natural Science Foundation(JQ19032), and the Qiu Shi Science & Technologies Foundation.

**Figure titles and legends**

**Figure 1.** The fossil fuel and cement $CO_2$ emissions distributions of GRACED in 2020. The value is given in the unit of kg of carbon per day per cell.

**Figure 2.** The range value of daily variations of total emissions in 2020.

**Figure 3.** Map of weekend minus weekday emissions in 2020.

**Figure 4.** Difference in daily average $CO_2$ emissions between 2020 and 2019 (2020 minus 2019).

**Figure 5.** The day of the year with (a) the maximum reduction, and (b) the maximum rebound of each grid in 2020 compared with 2019.

**Figure 6.** Sector share of $CO_2$ emissions in 2020(a-g). Difference in sector share of $CO_2$ emissions between 2020 and 2019(2020 minus 2019)(h-n).

**Figure 7.** The framework of top-down spatially gridding methodology.

**Table titles and legends**

**Table 1.** Correspondence between Carbon Monitor categories and GID/EDGAR categories.

**Table 1. Correspondence between Carbon Monitor categories and GID/EDGAR categories.**

| No. | Carbon Monitor categories | GID categories | EDGAR categories |
|---|---|---|---|
| 1 | Power | power | |
| 2 | Industry | industry | |
| 3 | Residential consumption | resident | |
| 4 | Ground transport | transport | |
| 5 | International aviation | | AIR Bunker oil for international transport |
| 6 | International shipping | | SEA Bunker oil for international transport |
| 7 | Domestic aviation | | AIR Bunker oil for domestic transport |

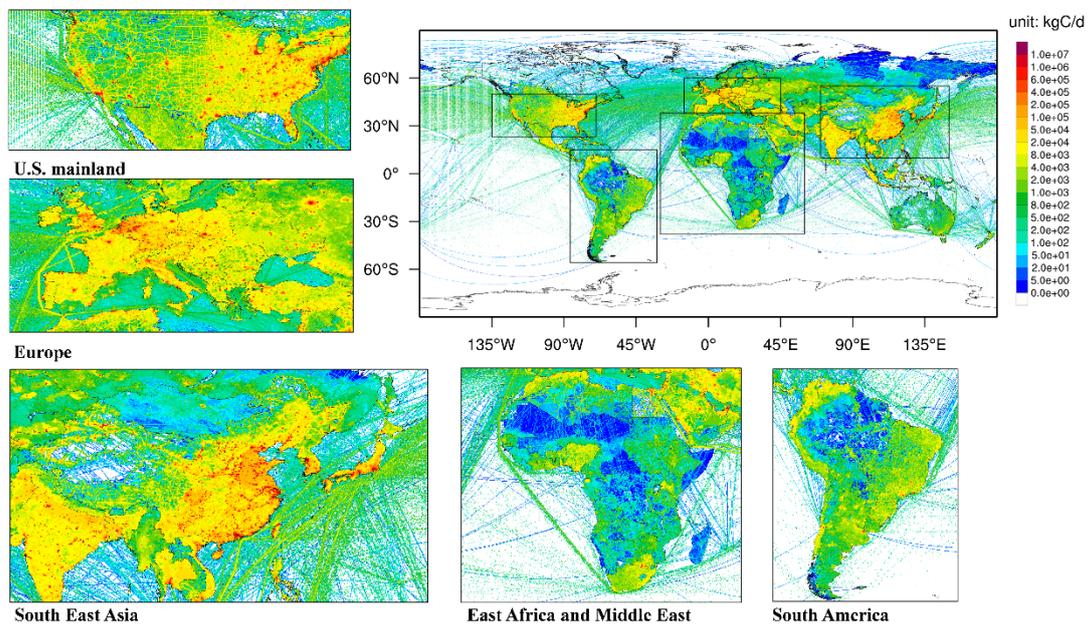

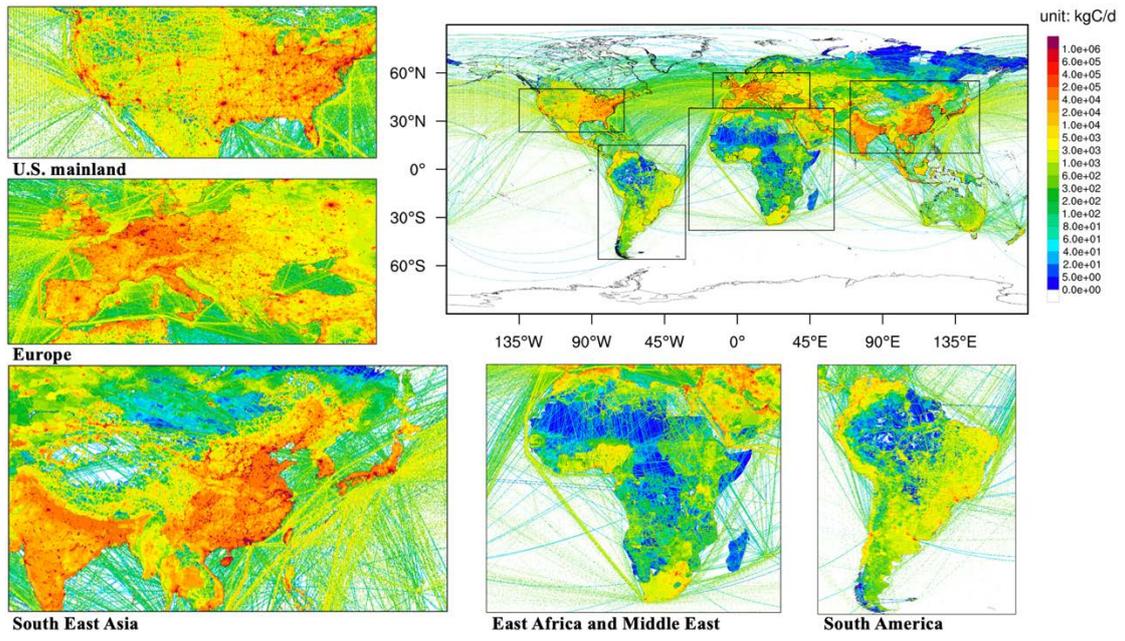
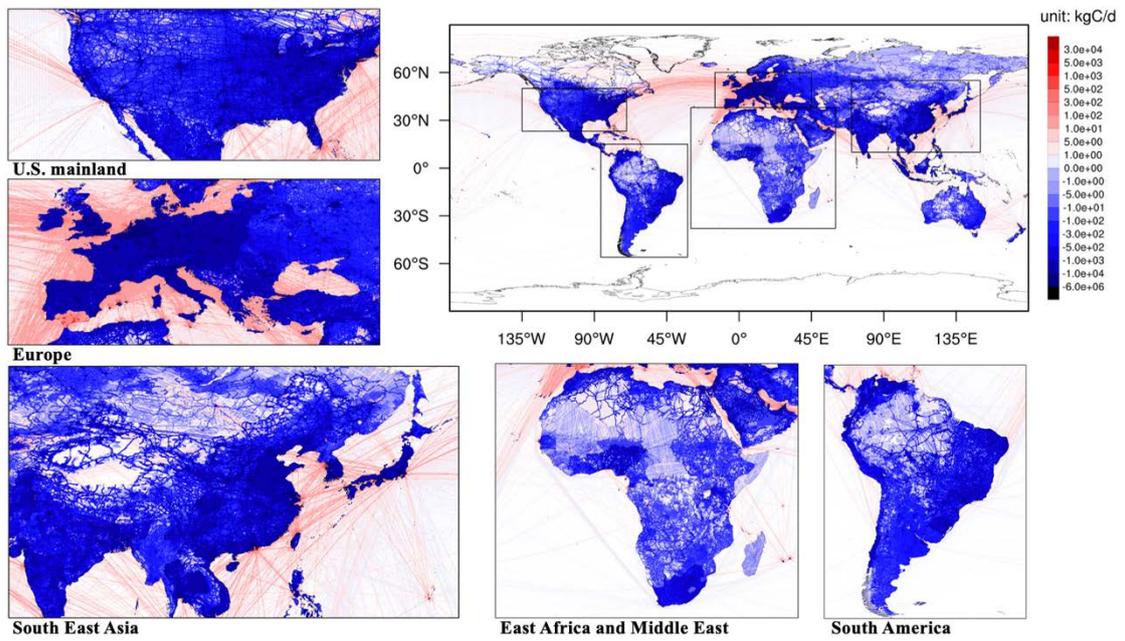

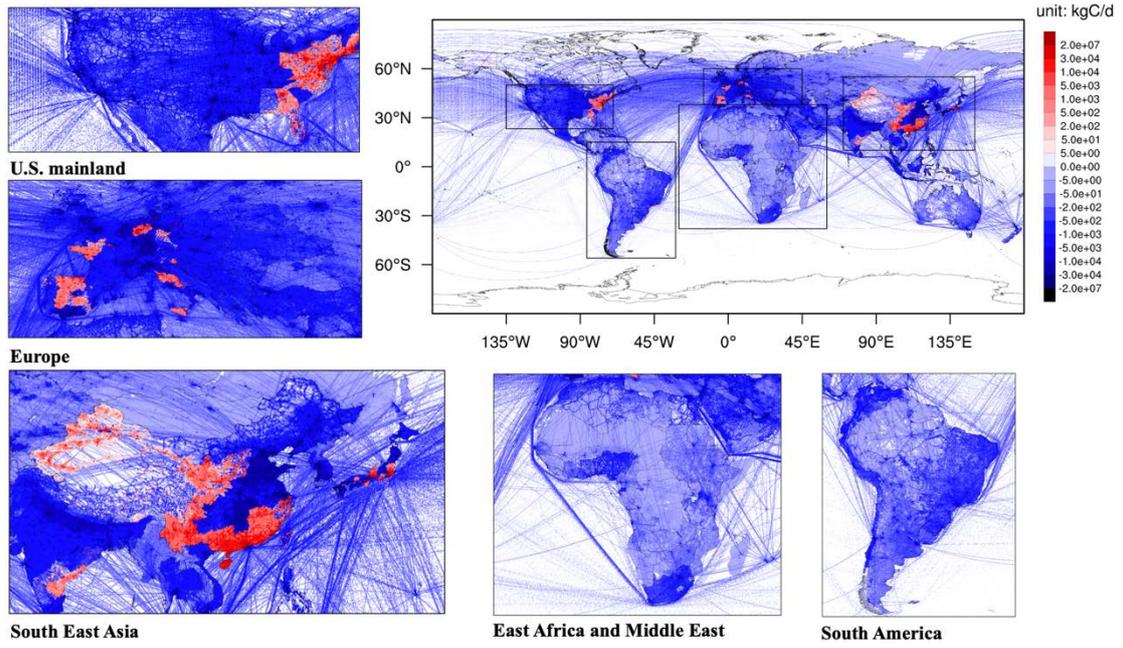

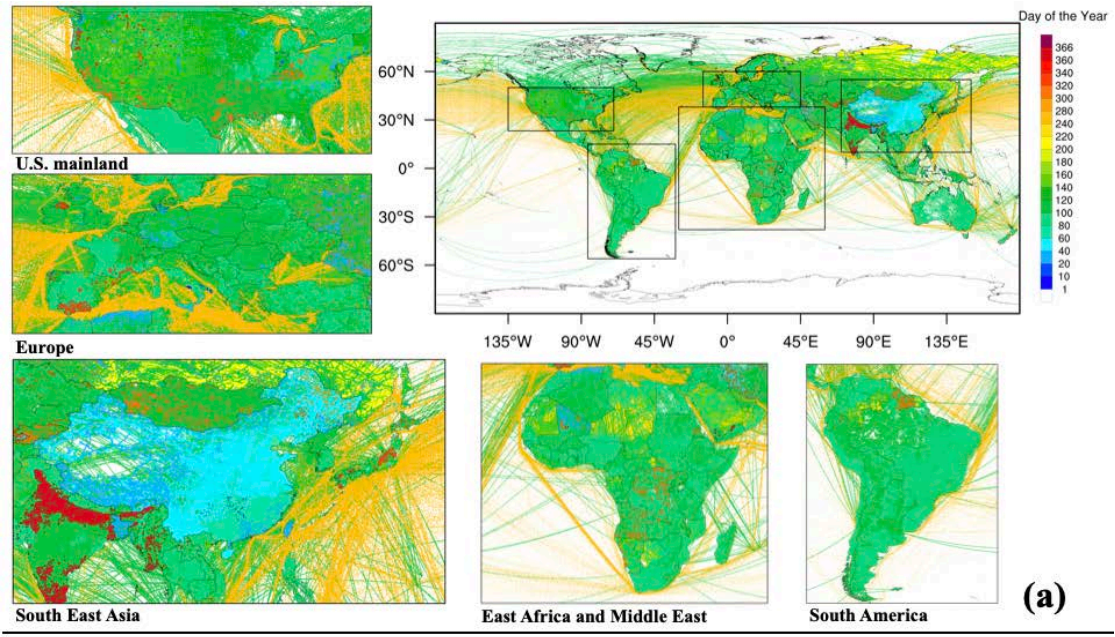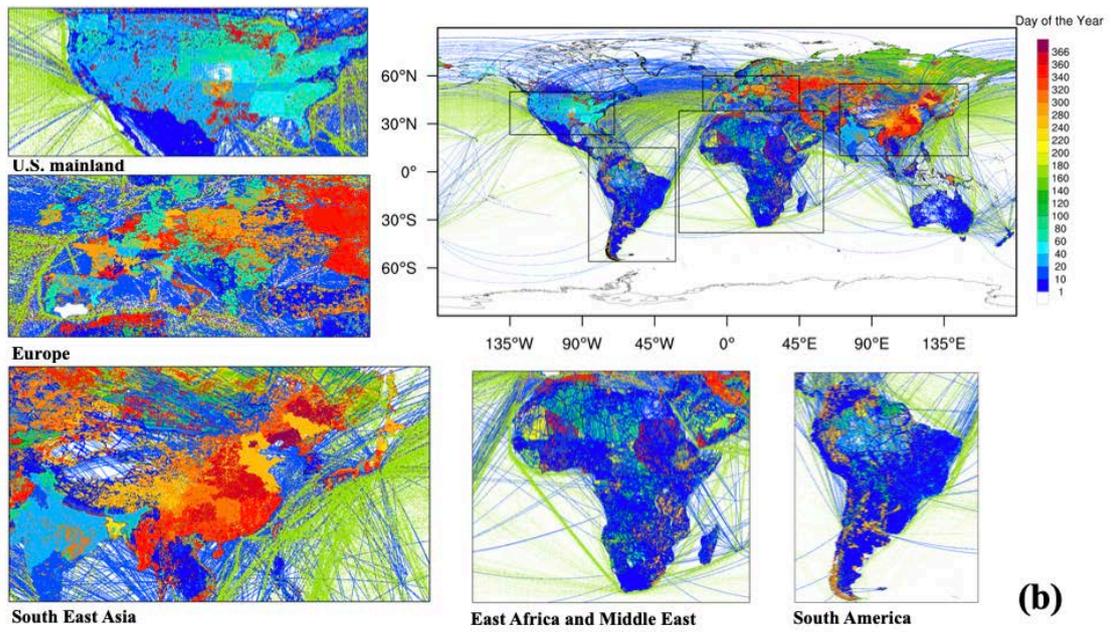

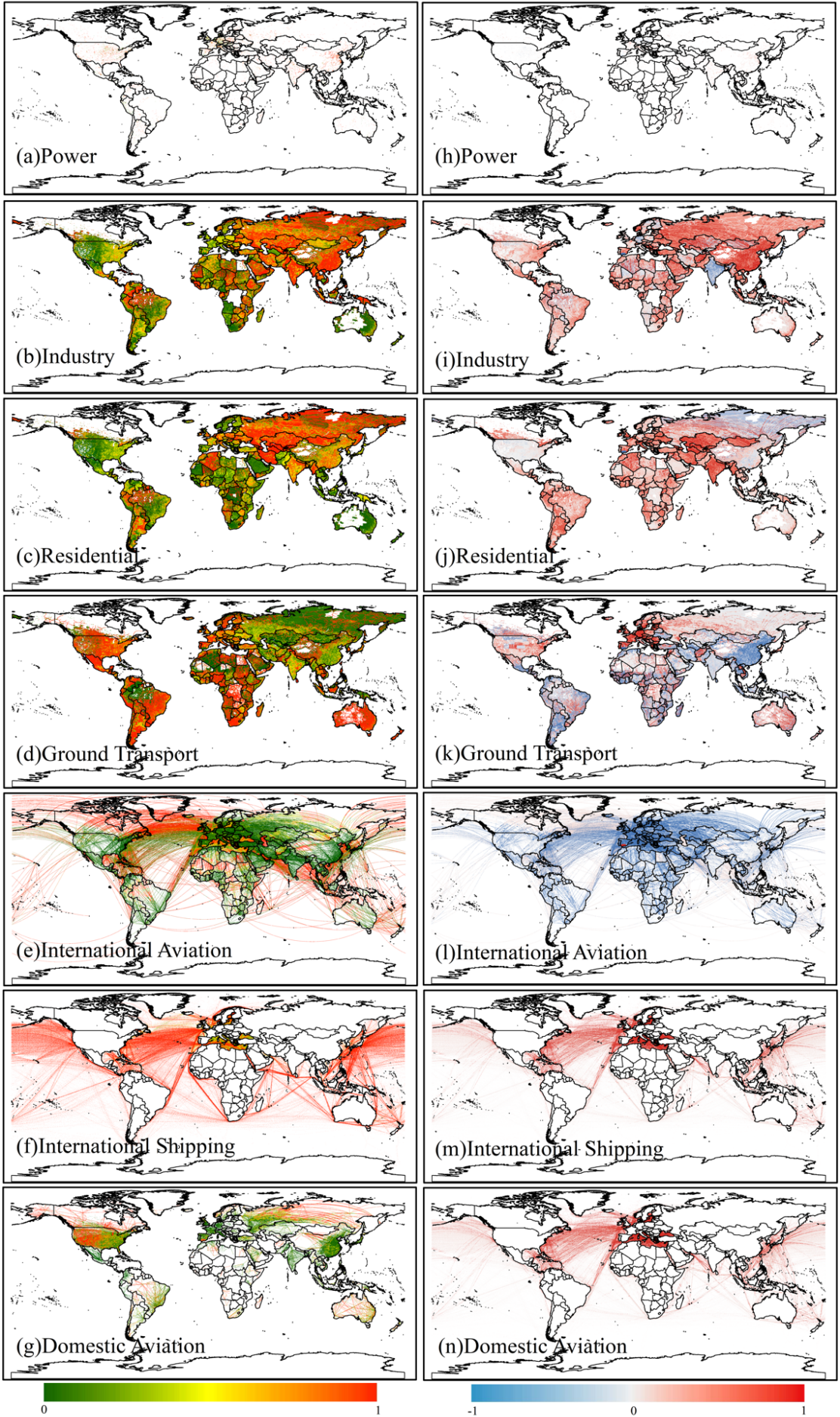

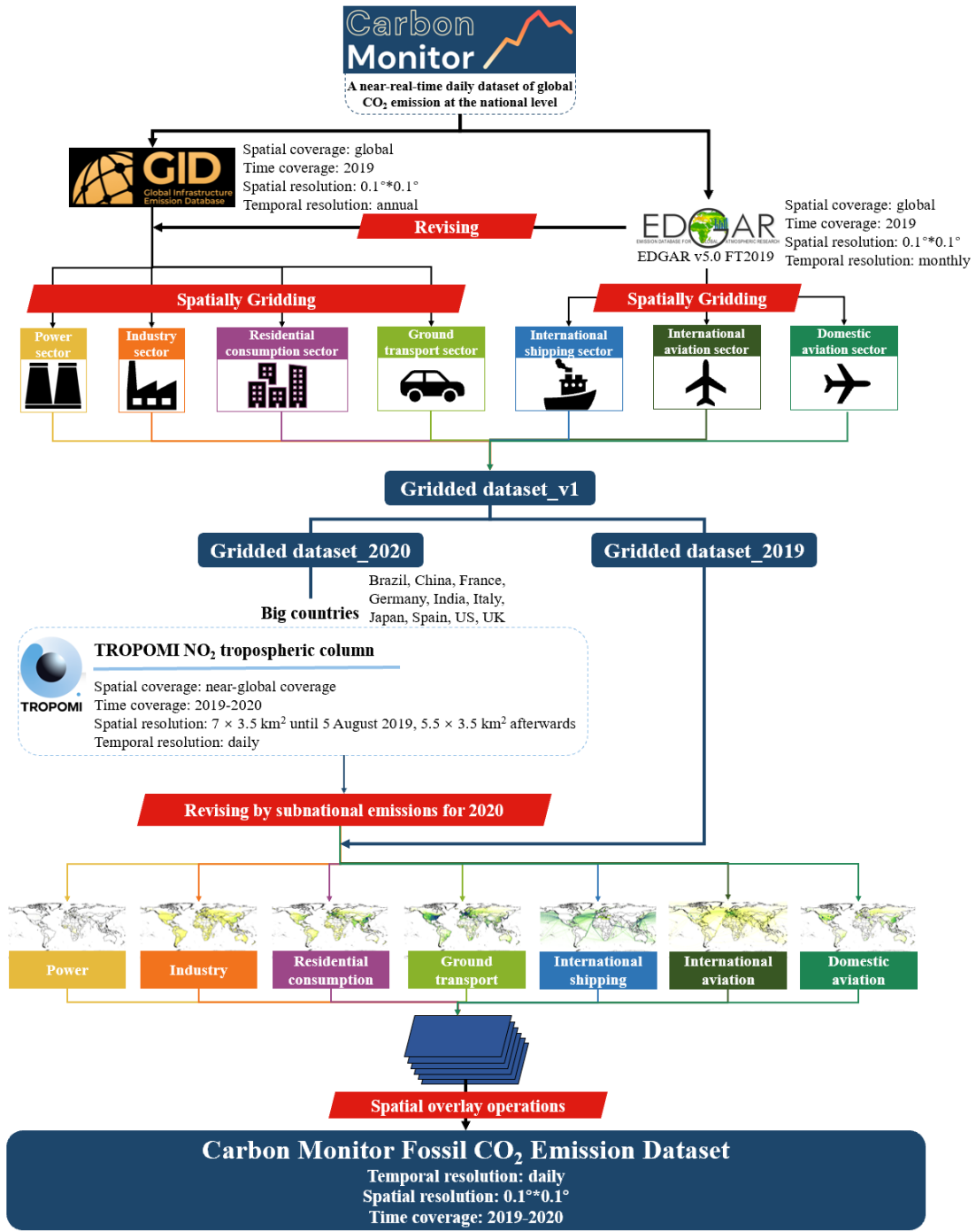